# Photonic radio frequency and microwave integration based on a multi-wavelength 49GHz Kerr microcomb source

Xingyuan Xu, Mengxi Tan, Jiayang Wu, Andreas Boes, *Member, IEEE*, Bill Corcoran, Thach G. Nguyen, Sai T. Chu, Brent E. Little, Roberto Morandotti, *Senior Member, IEEE*. Arnan Mitchell, *Member, IEEE*, and David J. Moss, *Fellow, IEEE*

*Abstract*—We demonstrate a photonic RF integrator based on an integrated soliton crystal micro-comb source. By multicasting and progressively delaying the input RF signal using a transversal structure, the input RF signal is integrated discretely. Up to 81 wavelengths are provided by the microcomb source, which enable a large integration time window of ~6.8 ns, together with a time resolution as fast as ~84 ps. We perform signal integration of a diverse range of input RF signals including Gaussian pulses with varying time widths, dual pulses with varying time intervals and a square waveform. The experimental results show good agreement with theory. These results verify our microcomb-based integrator as a competitive approach for RF signal integration with high performance and potentially lower cost and footprint.

## I. Introduction

Temporal integration is a basic function that has a wide range of applications in signal processing systems. In contrast to electrical integrators that are subject to the electronic bandwidth bottleneck, photonic techniques offer distinct advantages such as a broad bandwidth, strong immunity to electromagnetic interference, and low loss [1-3], thus holding great promise to overcome the limitations of their electrical counterparts.

Extensive effort has been made to achieve photonic integrators (see Table 1), such as those based on gratings [4-6], and micro-ring resonators (MRRs) [7-9]. These approaches achieve optical signal integration with a time feature as fast as 8 ps [7] and a large time-bandwidth product with high-Q resonant structures.

However, these approaches still face limitations for RF signal processing, as they are not reconfigurable in terms of temporal resolution and the length of the integration window, thus lacking the flexibility of processing RF signals with different bandwidths and varying target integration time windows. In addition, those approaches are specifically designed to process optical signals, thus they cannot directly process RF signals until being further equipped with electrical to optical interfaces.

Other approaches to photonic integrators based on transversal structures offer high reconfigurability and accuracy owing to the parallel scheme where each path can be controlled independently [10-12]. By tailoring the progressive delay step, RF signal integration with a reconfigurable operation bandwidth was achieved [10-12]. Yet these integrators still face limitations arising from the limited number of channels. To establish multiple wavelength channels, discrete laser arrays or electro-optical comb sources are generally employed, which suffer from a trade-off between the number of wavelengths and system complexity and ultimately lead to limited channel numbers, and in turn, time-bandwidth products.

Recently, a novel multi-wavelength source — integrated microcombs [13-19] — has attracted significant interest in RF photonic systems [20, 21]. These arise from optical parametric oscillation in ultra-high-Q monolithic MRRs and offer many advantages over traditional multi-wavelength sources, including a much higher number of wavelengths and a greatly reduced footprint and complexity for the system. A wide range of RF applications have been demonstrated based on microcombs, such as RF true time delays [22, 23], transversal signal processors [24-29], frequency conversion [30], phase-encoded signal generators [31], and channelizers [32, 33].

In this paper, we demonstrate a highly reconfigurable photonic RF integrator using an integrated soliton crystal micro-comb source [34, 35]. The input RF signal is multicast onto the flattened microcomb lines and progressively delayed via dispersion, and then summed upon detection to achieve temporal integration. The large number of wavelengths - up to 81 - offered by the microcomb enable a large integration time window of ~6.8 ns with a time resolution as fast as ~84 ps. A comb shaping system is developed to compensate for the nonuniform spectral output of the soliton crystal microcomb. We successfully tested the system with a range of different input signals. The experimental results match with theory well, verifying the performance and feasibility of our approach towards the realization of large-time-window photonic RF integration with potentially lower cost and footprint.

Table 1. Parameters for photonic RF integrators

| Method | Reconfigurability | Time window | Time resolution | Bandwidth | Time-bandwidth product |
|---|---|---|---|---|---|
| Fiber Bragg grating [4] | × | N.A. | 6ps | N.A. | N.A. |
| Active fiber Bragg grating [5] | × | ~5 ns | 50 ps | 20 GHz | 100 |
| Apodized uniform period fiber Bragg grating [6] | × | 60 ps | 2.5 ps | 400 GHz | 24 |
| Passive high-Q MRR [7] | × | 800 ps | 8 ps | 200 GHz | 100 |
| InP-InGaAsP MRR [8] | × | 6.33 ns | < 54 ps | >15 GHz | >117 |
| Passive MRR [9] | × | 12.5 ps | 1.9 ps | 400 GHz | 6.6 |
| Incoherent light source [10] | √ | 200 ps | 7 ps | >18 GHz | 28.5 |
| Multi-wavelength fiber laser [11] | √ | 1.24 ns | ~27 ps | 36.8 GHz | ~45.9 |
| Incoherent light source and fiber Bragg grating [12] | √ | 7 ns | 344 ps | 2.9 GHz | 20.3 |
| Microcomb (this work) | √ | 6.8 ns | 84 ps | 11.9 GHz | 81 |

## II. EXPERIMENTAL RESULTS

Figure 1 illustrates the operation principle of the photonic RF integrator. The integration process can be achieved via a discrete time-spectrum convolution operation between the RF input signal $f(t)$ and the flattened microcomb. With a delay step of $\Delta t$, the operation can be described as:

$$y(t) = \sum_{k=1}^{N} f(t + k \cdot \Delta t) \tag{1}$$

where $N$ is the total number of wavelength channels. After the replicas of $f(t)$ are delayed progressively and summed together, the integration of $f(t)$ can be achieved, with a time feature [10] of $\Delta t$ and a total integration time window of $T = N \times \Delta t$.

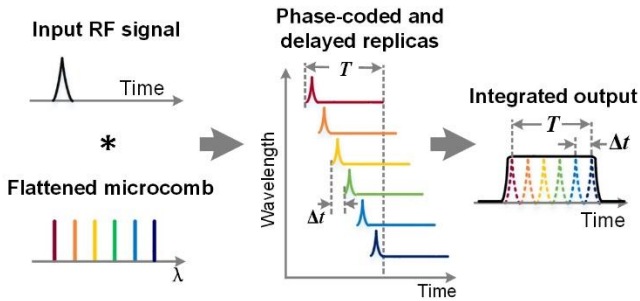

Figure 1. Schematic diagram of the photonic RF integration.

Figure 2 shows the experimental up of the photonic RF integrator using a microcomb source. The microcomb was generated by pumping a nonlinear high-Q MRR (Q factor > 1.5 million, free spectral range = ~0.4nm or ~48.9 GHz) with a continuous-wave (CW) laser. As the pump power and wavelength detuning were adjusted to provide sufficient parametric gain, soliton crystal microcombs were generated. The soliton crystal microcombs [21], leading to tightly packed solitons circulating in the MRR, were generated in our experiments due to a mode crossing at ~1552 nm of the MRR. The distinctive palm-like comb spectra (Fig. 3) is a result of the spectral interference between the circulating solitons. We then flattened the microcomb spectral lines in the C band with two stages of WaveShapers (Finisar 4000S). The input RF signal was imprinted onto the comb lines, generating replicas across all wavelength channels. The replicas were progressively delayed by a spool of standard single-mode fibre (~13km) and summed upon photodetection using a high-speed photodetector (Finisar, 40 GHz bandwidth).

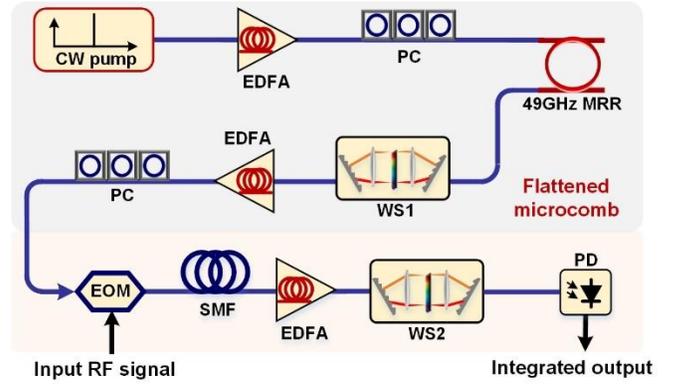

Figure 2. Experimental setup of the photonic RF integrator based on a micro-comb source. EDFA: erbium-doped fibre amplifier. PC: polarization controller. MRR: micro-ring resonator. WS: WaveShaper. EOM: Mach-Zehnder modulator. SMF: single mode fibre. PD: photodetector.

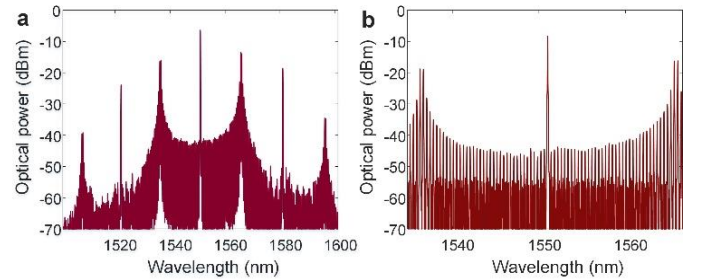

Figure 3. Optical spectra of the generated soliton crystal microcomb with spans of (a) 100 nm and (b) 30 nm.

The delay step between the adjacent wavelength channels $\Delta t$, or referred as the time feature of the integrator [10], was measured to be ~ 84 ps, which was determined by the dispersion and length of the fibre and the spectral spacing between the comb lines. We note that the fastest time feature is determined by the delay step of the wavelength channels, and thus in theory can become arbitrarily small by reducing the amount of dispersion, although with a tradeoff that the integration window also decreases proportionally.

The first WaveShaper (WS1) pre-flattened the optical spectrum of the microcomb to acquire a high link gain and a high signal-to-noise ratio. The second WaveShaper was employed for accurate comb shaping assisted by feedback control. The error signal was generated by reading and comparing the comb lines' power with the desired channel weights, which are all equal for the integrator. The measured optical spectrum after comb shaping is shown in Fig. 4.

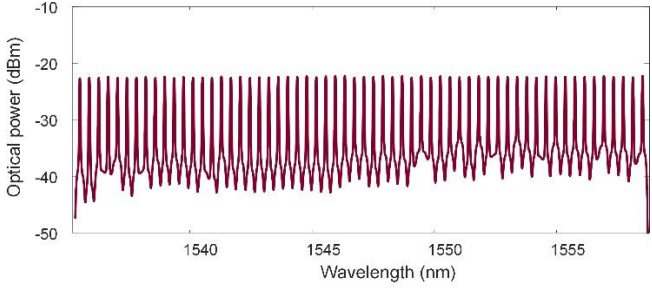

Figure 4. Optical spectrum of the shaped microcomb.

We selected 60 comb lines in the C-band for the integration, yielding a total integration time window ($T=N×\Delta t$) which reached up to 60 × 84 ps = 5.04 ns, as denoted by the yellow shaded region in Fig. 5. To verify the performance of our approach, we performed intervals of 1.52 ns and 3.06 ns, respectively. The measured results (red curves) clearly illustrate the performance of our integrator by exhibiting three distinct intensity steps in the integration waveforms. The left step corresponds to the integration of the first pulse while the middle step indicates the integration of both the two initial pulses, and the right step shows the integration of only the second pulse since it is beyond the integration window of the first pulse. Moreover, the performance of the integrator is further demonstrated by an input signal with a rectangular input waveform with its width equal to the integration window (5 ns). The measured integrated waveform exhibits a triangular shape that matches well with ideal integration results (gray curve).

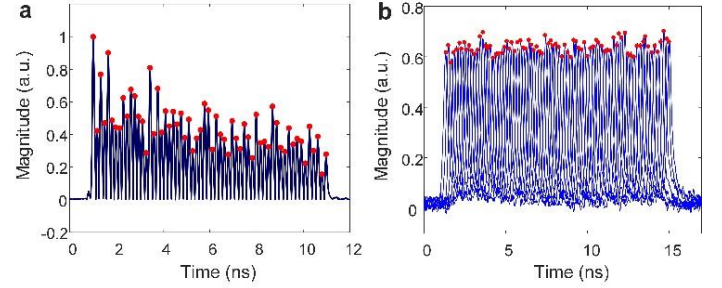

Figure 6. Measured impulse response of the integrator (a) after comb optical power shaping and (b) after impulse response shaping using a Gaussian RF input pulse.

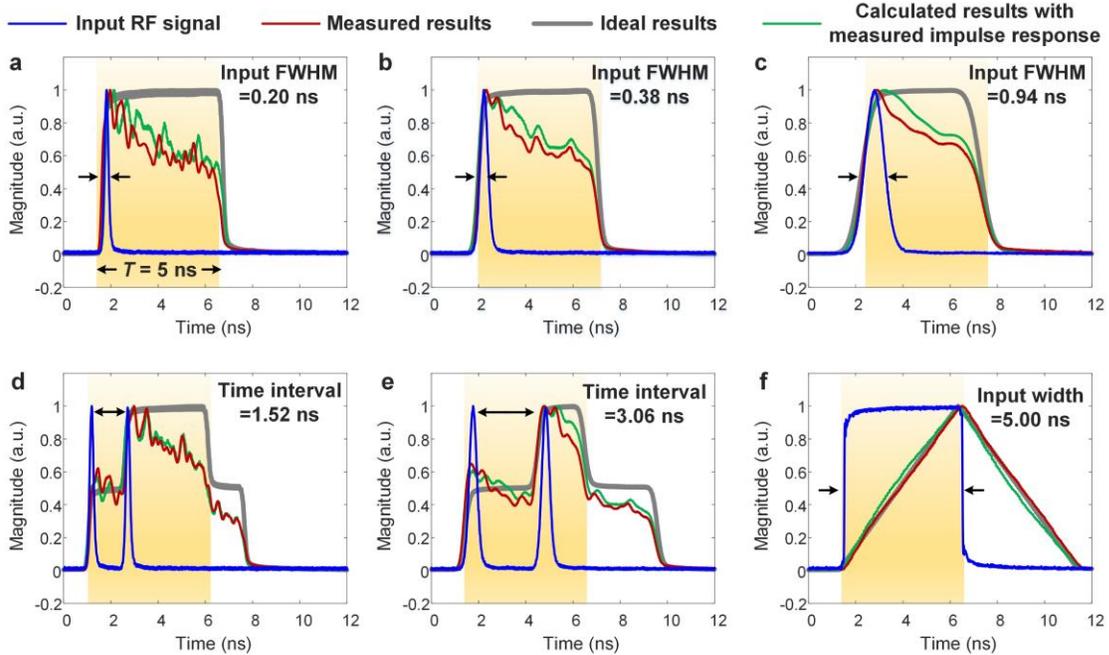

Figure 5. Experimental results of the microcomb-based RF integrator after comb optical power shaping for input (a-c) Gaussian pulses with FWHM of 0.20, 0.38 and 0.94 ns, (d-e) dual Gaussian pulses with time intervals of 1.52 and 30.6 ns, and (f) a triangular waveform with a width of 5.00 ns. The blue curves denote the input signal, the red curves denote the measured integration results, the gray curves denote the ideal integration results, and the green curves denote the integration results calculated with the measured impulse response of the system.

signal integration for different RF input signals. The red curves in Fig. 5(a-c) show the integration results for Gaussian pulses (blue curves) with a full width at half maximum (FWHM) varying from 0.20 ns to 0.94 ns, where the demonstrated integration window $T$ (~5 ns) matched well with theory (5.04 ns). Fig. 5(d-e) shows the integration results of dual Gaussian pulses with different time

## III. RESULT ANALYSIS AND OPTIMIZATION

As shown in Fig. 5, we note that there are residual discrepancies between the measured (red curves) and the ideal (grey curves). Considering that the optical power of the comb lines has been flattened well, we deduce that the errors were introduced by the non-ideal impulse response of the system, caused by the non-flat transmission response of the optical amplifier, the modulation and the photodetector across the wavelength channels. To verify this, we measured the impulse response of the system with a Gaussian pulse input. Considering that the time resolution of the system (~ 84 ps) was much smaller than the duration of the input pulse, we separated the wavelength channels into multiple subsets (each with a much larger spacing between the adjacent comb lines and thus obtained a temporal resolution larger than the input pulse duration), and measured their impulse response sequentially. Figure 6(a) shows the measured impulse response of the system, which was not flat in magnitude even when the comb lines were perfectly flattened. We used the measured impulse response and the input RF signal in Fig. 5 to calculate the corresponding integral output, with the results matching the experimentally measured integration well — verifying our deductions that the experimental errors were induced by the non-ideal impulse responses of the system.

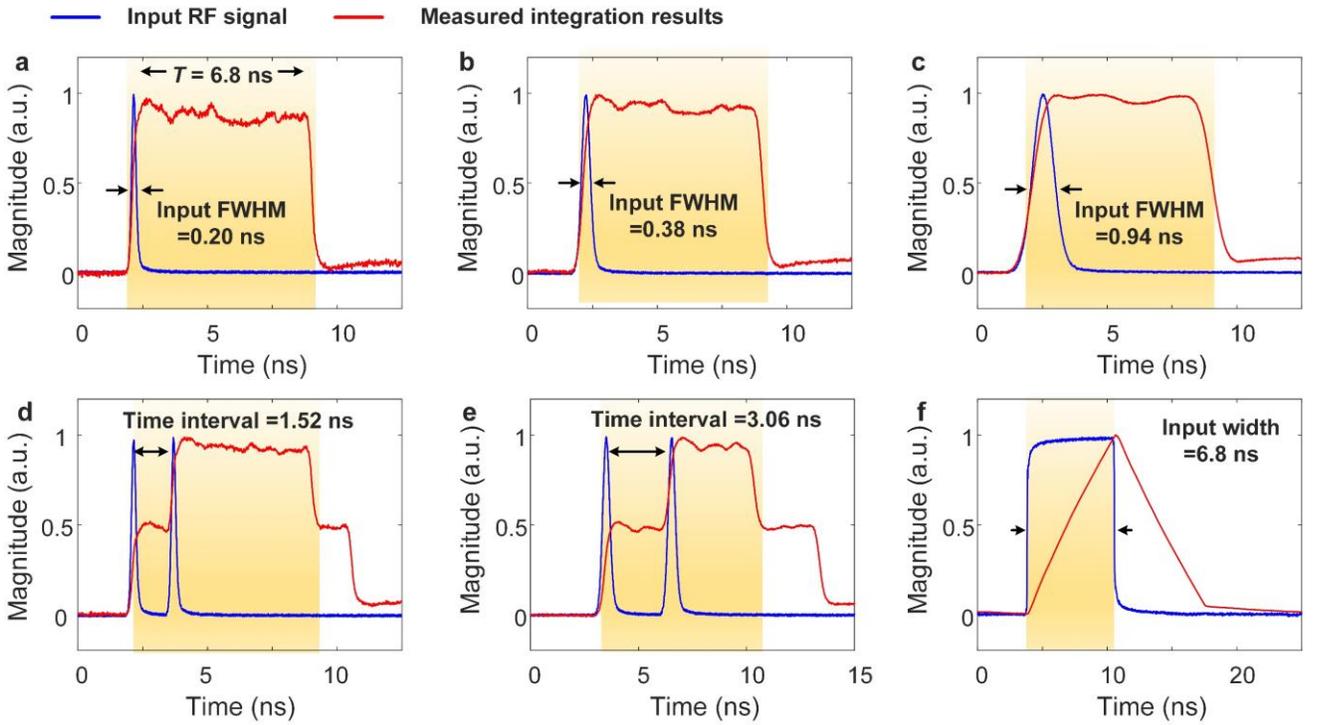

Figure 7. Experimental results of the microcomb-based RF integrator after impulse response shaping for input (a-c) Gaussian pulses with FWHM of 0.20, 0.38 and 0.94 ns, (d-e) dual Gaussian pulses with time intervals of 1.52 and 3.06 ns, and (f) a triangular waveform with a width of 5.00 ns. The blue curves denote the input signal, the red curves denote the measured integration results.

In order to reduce these errors, we developed a more accurate comb shaping approach, where the error signal of the feedback loop was generated directly by the measured impulse response, instead of the optical power of the comb lines. The resulting flattened impulse response is shown in Fig. 6(b), which is much closer to the ideal impulse response than Fig. 6(a).

We then performed integration with the same RF inputs as for previous measurements, the results of which are shown in Fig. 7. Note that during this measurement, 81 wavelength channels were enabled by the impulse response shaping process, so that the integration time window ($T = N \times \Delta t$) increased to 81 × 84ps = 6.804 ns, resulting in an operation bandwidth of 1/84ps = 11.9 GHz and a time-bandwidth product of 6.804ns × 11.9GHz = ~81 (approximately equal to the number of channels $N$). The measured integrated results (red curves, Fig. 7) show significantly better agreement with theory, indicating the success of the impulse response shaping method, and the feasibility of our approach to photonic RF integration based on microcombs.

## IV. Conclusions

In conclusion, we demonstrate a photonic RF integrator using an integrated soliton crystal micro-comb source. Through broadcast and delay processes employing 81 wavelength channels generated by the microcomb source, discrete temporal integration of RF signals is achieved. A large integration time window of 6.8 ns is demonstrated, together with a time feature as fast as 84 ps. An impulse response shaping approach was developed to compensate for the non-flat optical transmission response of the system to guarantee uniform channels weights for the integrator. Different input signals were successfully integrated. These results verify that our RF integrator is a competitive approach towards integrating high-speed photonic RF signals with high performance, flexibility, and potentially reduced cost and footprint.

## Acknowledgment

This work was supported by the Australian Research Council Discovery Projects Program (No. DP150104327). RM acknowledges support by the Natural Sciences and Engineering Research Council of Canada (NSERC) through the Strategic, Discovery and Acceleration Grants Schemes, by the MESI PSR-SIIRI Initiative in Quebec, and by the Canada Research Chair Program. Brent E. Little was supported by the Strategic Priority Research Program of the Chinese Academy of Sciences, Grant No. XDB24030000.

## References

[1] J. Capmany, and D. Novak, "Microwave Photonics combines two worlds," *Nature Photonics*, vol. 1, pp. 319-330, Jun. 2007.

[2] J. Yao, "Microwave Photonics," *Journal of Lightwave Technology*, vol. 27, no. 3, pp. 314-335, Feb. 2009.

[3] R. C. Williamson, R. D. Esman, "RF Photonics," *Journal of Lightwave Technology*, vol. 26, no. 9, pp. 1145-1153, May 2008.

[4] Y. Park, T.-J. Ahn, Y. Dai, J. Yao, and J. Azaña, "All-optical temporal integration of ultrafast pulse waveforms," *Optics Express,* vol. 16, no. 22, pp. 17817-17825, 2008/10/27. 2008.

[5] R. Slavík, Y. Park, N. Ayotte, S. Doucet, T.-J. Ahn, S. LaRochelle, and J. Azaña, "Photonic temporal integrator for all-optical computing," *Optics Express,* vol. 16, no. 22, pp. 18202-18214, 2008/10/27. 2008.

[6] M. H. Asghari, Y. Park, and J. Azaña, "New design for photonic temporal integration with combined high processing speed and long operation time window," *Optics Express,* vol. 19, no. 2, pp. 425-435, 2011/01/17. 2011.

[7] M. Ferrera, Y. Park, L. Razzari, B. E. Little, S. T. Chu, R. Morandotti, D. J. Moss, and J. Azaña, "On-chip CMOS-compatible all-optical integrator," *Nature Communications,* vol. 1, no. 1, pp. 29, 2010/06/15. 2010.

[8] W. Liu, M. Li, R. S. Guzzon, E. J. Norberg, J. S. Parker, L. A. Coldren, and J. Yao, "A Photonic Temporal Integrator With an Ultra-Long Integration Time Window Based on an InP-InGaAsP Integrated Ring Resonator," *Journal of Lightwave Technology,* vol. 32, no. 20, pp. 3654-3659. 2014.

[9] M. Ferrera, Y. Park, L. Razzari, B. E. Little, S. T. Chu, R. Morandotti, D. J. Moss, and J. Azaña, "All-optical 1st and 2nd order integration on a chip," *Optics Express,* vol. 19, no. 23, pp. 23153-23161, 2011/11/07. 2011.

[10] Y. Park, and J. Azaña, "Ultrafast photonic intensity integrator," *Optics Letters,* vol. 34, no. 8, pp. 1156-1158, 2009/04/15. 2009.

[11] A. Malacarne, R. Ashrafi, M. Li, S. LaRochelle, J. Yao, and J. Azaña, "Single-shot photonic time-intensity integration based on a time-spectrum convolution system," *Optics Letters,* vol. 37, no. 8, pp. 1355-1357, 2012/04/15. 2012.

[12] J. Zhang, and J. Yao, "Microwave photonic integrator based on a multichannel fiber Bragg grating," *Optics Letters,* vol. 41, no. 2, pp. 273-276, 2016/01/15. 2016.

[13] T. J. Kippenberg, A. L. Gaeta, M. Lipson, and M. L. Gorodetsky, "Dissipative Kerr solitons in optical microresonators," *Science,* vol. 361, no. 6402, pp. 8083. 2018.

[14] D. J. Moss, R. Morandotti, A. L. Gaeta, and M. Lipson, "New CMOS-compatible platforms based on silicon nitride and Hydex for nonlinear optics," *Nature Photonics,* vol. 7, no. 8, pp. 597-607, Aug. 2013.

[15] A. L. Gaeta, M. Lipson, and T. J. Kippenberg, "Photonic-chip-based frequency combs," *Nature Photonics,* vol. 13, no. 3, pp. 158-169, Mar. 2019.

[16] A. Pasquazi, M. Peccianti, L. Razzari, D. J. Moss, S. Coen, M. Erkintalo, Y. K. Chembo, T. Hansson, S. Wabnitz, P. Del'Haye, X. X. Xue, A. M. Weiner, and R. Morandotti, "Micro-combs: A novel generation of optical sources," *Physics Reports-Review Section of Physics Letters,* vol. 729, pp. 1-81, Jan 27. 2018.

[17] M.-G. Suh, and K. J. Vahala, "Soliton microcomb range measurement," *Science,* vol. 359, no. 6378, pp. 884-887. 2018.

[18] W. Liang, D. Eliyahu, V. S. Ilchenko, A. A. Savchenkov, A. B. Matsko, D. Seidel, and L. Maleki, "High spectral purity Kerr frequency comb radio frequency photonic oscillator," *Nature Communications,* vol. 6, pp. 7957. 2015.

[19] D. T. Spencer, T. Drake, T. C. Briles, J. Stone, L. C. Sinclair, C. Fredrick, Q. Li, D. Westly, B. R. Ilic, and A. Bluestone, "An optical-frequency synthesizer using integrated photonics," *Nature,* vol. 557, no. 7703, pp. 81-85. 2018.

[20] J. Wu, X. Xu, T. G. Nguyen, S. T. Chu, B. E. Little, R. Morandotti, A. Mitchell, and D. J. Moss, "RF Photonics: An Optical Microcombs' Perspective," *IEEE Journal of Selected Topics in Quantum Electronics,* vol. 24, no. 4, pp. 1-20. 2018.

[21] X. Xu, M. Tan, J. Wu, R. Morandotti, A. Mitchell, and D. J. Moss, "Microcomb-based photonic RF signal processing," *IEEE Photonics Technology Letters,* vol. 31, no. 23, pp. 1854-1857. 2019.

[22] X. Xu, J. Wu, T. G. Nguyen, T. Moein, S. T. Chu, B. E. Little, R. Morandotti, A. Mitchell, and D. J. Moss, "Photonic microwave true time delays for phased array antennas using a 49 GHz FSR integrated optical micro-comb source [Invited]," *Photonics Research,* vol. 6, no. 5, pp. B30-B36, May 1. 2018.

[23] X. Xue, Y. Xuan, C. Bao, S. Li, X. Zheng, B. Zhou, M. Qi, and A. M. Weiner, "Microcomb-Based True-Time-Delay Network for Microwave Beamforming With Arbitrary Beam Pattern Control," *Journal of Lightwave Technology,* vol. 36, no. 12, pp. 2312-2321, Jun. 2018.

[24] X. Xu, J. Wu, T. G. Nguyen, M. Shoeiby, S. T. Chu, B. E. Little, R. Morandotti, A. Mitchell, and D. J. Moss, "Advanced RF and microwave functions based on an integrated optical frequency comb source," *Optics Express,* vol. 26, no. 3, pp. 2569-2583, Feb 5. 2018.

[25] X. Xu, M. Tan, J. Wu, T. G. Nguyen, S. T. Chu, B. E. Little, R. Morandotti, A. Mitchell, and D. J. Moss, "Advanced Adaptive Photonic RF Filters with 80 Taps Based on an Integrated Optical Micro-Comb Source," *Journal of Lightwave Technology,* vol. 37, no. 4, pp. 1288-1295, Feb. 2019.

[26] X. X. Xue, Y. Xuan, H. J. Kim, J. Wang, D. E. Leaird, M. H. Qi, and A. M. Weiner, "Programmable Single-Bandpass Photonic RF Filter Based on Kerr Comb from a Microring," *Journal of Lightwave Technology,* vol. 32, no. 20, pp. 3557-3565, Oct 15. 2014.

[27] X. Xu, J. Wu, M. Shoeiby, T. G. Nguyen, S. T. Chu, B. E. Little, R. Morandotti, A. Mitchell, and D. J. Moss, "Reconfigurable broadband microwave photonic intensity differentiator based on an integrated optical frequency comb source," *APL Photonics,* vol. 2, no. 9, Sep. 2017.

[28] M. Tan, X. Xu, B. Corcoran, J. Wu, A. Boes, T. G. Nguyen, S. T. Chu, B. E. Little, R. Morandotti, A. Mitchell, and D. J. Moss, "Microwave and RF Photonic Fractional Hilbert Transformer Based on a 50 GHz Kerr Micro-Comb," *Journal of Lightwave Technology,* vol. 37, no. 24, pp. 6097-6104. 2019.

[29] M. Tan, X. Xu, B. Corcoran, J. Wu, A. Boes, T. G. Nguyen, S. T. Chu, B. E. Little, R. Morandotti, A. Mitchell, and D. J. Moss, "RF and microwave fractional differentiator based on photonics," IEEE Transactions on Circuits and Systems II: Express Briefs, Early Access. 2020. DOI: 10.1109/TCSII.2020.2965158

[30] X. Xu, J. Wu, M. Tan, T. G. Nguyen, S. Chu, B. Little, R. Morandotti, A. Mitchell, and D. J. Moss, "Micro-comb based photonic local oscillator for broadband microwave frequency conversion," *Journal of Lightwave Technology*, vol. 38, no. 2, pp. 332-338. 2020.

[31] X. Xu, M. Tan, J. Wu, A. Boes, B. Corcoran, T. G. Nguyen, S. T. Chu, B. Little, R. Morandotti, and A. Mitchell, "Photonic RF phase-encoded signal generation with a microcomb source," *Journal of Lightwave Technology*, Early Access. 2019. DOI: 10.1109/JLT.2019.2958564

[32] X. Xu, M. Tan, J. Wu, T. G. Nguyen, S. T. Chu, B. E. Little, R. Morandotti, A. Mitchell, and D. J. Moss, "High performance RF filters via bandwidth scaling with Kerr micro-combs," *APL Photonics,* vol. 4, no. 2, pp. 026102. 2019.

[33] X. Xu, J. Wu, T. G. Nguyen, S. Chu, B. Little, A. Mitchell, R. Morandotti, and D. J. Moss, "Broadband RF Channelizer based on an Integrated Optical Frequency Kerr Comb Source," *Journal of Lightwave Technology,* vol. 36, no. 19, pp. 7. 2018.

[34] D. C. Cole, E. S. Lamb, P. Del'Haye, S. A. Diddams, and S. B. Papp, "Soliton crystals in Kerr resonators," *Nature Photonics,* vol. 11, no. 10, pp. 671. 2017.

[35] W. Wang, Z. Lu, W. Zhang, S. T. Chu, B. E. Little, L. Wang, X. Xie, M. Liu, Q. Yang, and L. Wang, "Robust soliton crystals in a thermally controlled microresonator," *Optics Letters,* vol. 43, no. 9, pp. 2002-2005. 2018.